\documentclass{ws-ijmpe}

\newcommand{\beq}{\begin{equation}}
\newcommand{\eeq}{\end{equation}}
\newcommand{\beqn}{\begin{eqnarray}}
\newcommand{\eeqn}{\end{eqnarray}}
\newcommand{\btab}{\begin{tabular}}
\newcommand{\etab}{\end{tabular}}

\begin{document}

\markboth{J.~Li, B.~Y.~Sun, and J.~Meng}{Pairing Properties of
Symmetric Nuclear Matter in Relativistic Mean Field Theory}

\catchline{}{}{}{}{}

\title{PAIRING PROPERTIES OF SYMMETRIC NUCLEAR MATTER\\IN RELATIVISTIC MEAN FIELD THEORY}

\author{\footnotesize J.~Li}

\address{School of Phyics, and State Key Laboratory of Nuclear Physics and Technology, Peking University\\
Beijing, 100871, P. R. China\\
Dipartimento di Fisica, Universit\`a degli Studi and INFN Sez. di Milano\\
Milano, 20133, Italy}

\author{B.~Y.~Sun}

\address{School of Phyics, and State Key Laboratory of Nuclear Physics and Technology, Peking University\\
Beijing, 100871, P. R. China}

\author{J.~Meng}

\address{School of Phyics, and State Key Laboratory of Nuclear Physics and Technology, Peking University\\
Beijing, 100871, P. R. China\\
Institute of Theoretical Physics, Chinese Academy of Sciences\\
Beijing, 100080, P. R. China\\
Center of Theoretical Nuclear Physics, National Laboratory of Heavy Ion Accelerator\\
Lanzhou, 730000, P. R. China\\
Department of Physics, University of Stellenbosch\\
Stellenbosch, South Africa\\
mengj@pku.edu.cn}

\maketitle

\begin{history}
\received{(received date)}
\revised{(revised date)}
\end{history}

\begin{abstract}
The properties of pairing correlations in symmetric nuclear matter
are studied in the relativistic mean field (RMF) theory with the
effective interaction PK1. Considering well-known problem that the
pairing gap at Fermi surface calculated with RMF effective
interactions are three times larger than that with Gogny force, an
effective factor in the particle-particle channel is introduced. For
the RMF calculation with PK1, an effective factor 0.76 give a
maximum pairing gap 3.2 MeV at Fermi momentum 0.9 fm$^{-1}$, which
are consistent with the result with Gogny force.
\end{abstract}

\section{Introduction}

The mean field theory, including non-relativistic mean field theory
with effective nucleon-nucleon interactions such as Skyrme or Gogny,
and the relativistic mean field (RMF) theory, has received wide
attention due to its successful descriptions of lots of nuclear
phenomena during the past years. In the framework of the RMF
theory,\cite{wale74} the nucleons interact via the exchanges of
mesons and photons. The representations with large scalar and vector
fields in nuclei provide simpler and more efficient descriptions
than non-relativistic approaches that hide these scales. In these
sense, the RMF theory is more fundamental. With a limited number of
free parameters, i.e. meson masses and meson-nucleon coupling
constants, the RMF theory has proved to be successful in
quantitatively describing the properties of nuclear matter and
neutron stars,\cite{glen97} nuclei near the valley of
stability,\cite{serto1986,reinhard1989,ring96} and exotic nuclei
with large neutron or proton excess with proper treatment of the
pairing correlations and continuum
effects.\cite{Meng:2006,Meng:1996,Meng:1998a,Meng:1998PRL} It gives
naturally the spin-orbit potential, the origin of the pseudo-spin
symmetry\cite{Arima:1969,Hecht:1969} as a relativistic
symmetry,\cite{Ginocchio97,Meng98r,Meng99prc,ChenTCCPL} and the spin
symmetry in the anti-nucleon spectrum.\cite{Zhou03prl}

Since 1950's, a large number of striking experimental facts, such as
the  binding energy difference between even-even and odd-even nuclei
and a systemic reduction of the moments of inertia of even-even
nuclei compared with their neighboring odd-even nuclei in deformed
nuclei, suggest the existence of the superfluid phenomena in these
systems.\cite{bohr1958pr} In astrophysics, the origin of pulsar
glitches,\cite{anderson1975n} which are sudden discontinuities in
the spin-down of pulsars, can also be understood via the
superfluidity in the inner crust of these stars. All of these
phenomena suggest that the pairing correlations play an important
role in theoretical study on the properties of nuclei structure and
nuclear matter.

The first relativistic study of superfluidity in infinite nuclear
matter was done by Kucharek and Ring in 1991.\cite{kucharek1991zpa}
They studied the pairing correlations in symmetric nuclear matter
using the relativistic Hartree-Bogoliubov (RHB) method with an
one-boson-exchange (OBE) potential in the particle-particle channel.
However the resulting pairing gap at the Fermi surface are about
three times larger than that with the Gogny
force.\cite{decharge1980prc} Therefore, the effective pairing
interaction used in RHB calculation are either the finite range
Gogny force or the Skyrme type zero-range force. In particular, with
the Skyrme type zero-range force, lots of achievements have been
made to describe the nuclei far away from the line of
$\beta$-stability with proper treatment of the pairing correlations
and continuum
effects.\cite{Meng:2006,Meng:1996,Meng:1998a,Meng:1998PRL}

However, it is still an open problem how the same nucleons
interaction via the exchanges of mesons and photons in the Hartree
channel can be used in the particle-particle channel as well. In
fact, the large pairing gap in RHB calculation with an OBE potential
in the particle-particle channel comes from the behavior of the
pairing matrix elements at large momenta.\cite{serra01} The various
effective forces in RMF models are adjusted for mean-field
calculations in Hartree channel only, i.e., only for momenta below
the Fermi momentum, thus a realistic particle-particle interaction
can have very different behavior at high momenta. Therefore, in
order to get reasonable values for the pairing gap, one can use a
suitable value of cutoff in the momentum space in the relativistic
mean field
calculations,\cite{ring96,serra01,guimaraes1996prc,matsuzaki1998prc}
or consider various effects, such as the medium polarization, the
in-medium meson mass decrease, and the mesons nonlinear terms to
reduce the pairing gap in nuclear
matter.\cite{chenjs2004plb,matsuzaki,sugimoto}

In this paper, the properties of pairing correlations in symmetric
nuclear matter will be studied in the RMF theory with the newly
developed effective interaction PK1.\cite{longwh2004prc} In order to
solve the well-known problem that the pairing gap at Fermi surface
calculated with RMF effective interactions are three times larger
than that with Gogny force, an effective factor in the
particle-particle channel will be introduced. In section II, a brief
description of the RMF theory and RHB theory for nuclear matter is
presented. The results and discussions are given in the following
section. Finally in the last section,  a brief summary is given.

\section{ Theoretical Framework}
\subsection{Relativistic Mean Field Theory}

A general review of the RMF theory and its application in nuclear
physics could be found in
Refs.~[\refcite{serto1986,reinhard1989,ring96}]. Here a brief review
of RMF theory for nuclear matter is given. The start point of RMF
theory is an effective Lagrangian density with the nucleons
interacting via the exchange of various mesons and the photon: \beqn
\label{lagrangian} {\cal L} &=&
    \bar{\psi}
    \left[
        i{\gamma^\mu}{\partial_\mu} - m - g_\sigma\sigma
        - g_\omega\gamma^\mu\omega_\mu
        - g_\rho\gamma^\mu\vec{\tau}\cdot\vec{\rho}_\mu
        - e\gamma^\mu\frac{1 - \tau_3}{2}A_\mu
    \right]
    \psi\nonumber\\
&&
    + \frac{1}{2}\partial^\mu\sigma\partial_\mu\sigma
    - \frac{1}{2}m_\sigma^2\sigma^2 - U(\sigma)
    - \frac{1}{4}\Omega^{\mu\nu}\Omega_{\mu\nu}
    + \frac{1}{2}m_\omega^2\omega^\mu\omega_\mu
    + U(\omega)\nonumber\\
&&
    - \frac{1}{4}\vec{R}^{\mu\nu}\vec{R}_{\mu\nu}
    + \frac{1}{2}m_\rho^2\vec{\rho}^\mu\cdot\vec{\rho}_\mu
    - \frac{1}{4}A^{\mu\nu}A_{\mu\nu}.
\eeqn The Dirac spinor $\psi$ denotes the nucleon with mass $m$. The
isoscalar scalar $\sigma$- and isoscalar vector $\omega$-mesons
offer medium-range attractive and short-range repulsive interactions
respectively, and the isovector vector $\rho$-meson provides the
necessary isospin asymmetry. Their masses are denoted by $m_\sigma$,
$m_\omega$ and $m_\rho$. $g_\sigma$, $g_\omega$ and $g_\rho$ are the
corresponding meson-nucleon coupling constants. $\tau$ is the
isospin of nucleon, and $\tau_3$ is its three-component. The
nonlinear $\sigma$ and $\omega$ self-interactions $U(\sigma)$ and
$U(\omega)$ are respectively denoted as,
 \beqn
    U(\sigma) &=& \frac{1}{3}g_2\sigma^3 + \frac{1}{4}g_3\sigma^4, ~~~
    U(\omega) ~=~ \frac{1}{4}c_3(\omega^\mu\omega_\mu)^2,
 \eeqn
with the self-coupling constants $g_2$, $g_3$ and $c_3$. The field
tensors $\Omega_{\mu\nu}$, $\vec{R}_{\mu\nu}$, and $A_{\mu\nu}$ are
as the followings,
 \beqn
    \Omega_{\mu\nu}  &=& \partial_\mu\omega_\nu - \partial_\nu\omega_\mu,~~
    \vec{R}_{\mu\nu} ~=~ \partial_\mu\vec{\rho}_\nu - \partial_\nu\vec{\rho}_\mu,~~
    A_{\mu\nu}       ~=~ \partial_\mu A_\nu - \partial_\nu A_\mu.
 \eeqn

The classical variation principle gives the following equations of
motion,
 \beqn
   \left[
      i\gamma^\mu\partial_\mu - m - g_\sigma\sigma -  g_\omega\gamma^\mu\omega_\mu
      - g_\rho\gamma^\mu\vec{\tau}\cdot\vec{\rho}_\mu - e\gamma^\mu\frac{1 - \tau_3}{2}A_\mu
   \right]\psi
&=& 0,
 \eeqn
 for the nucleon spinors and
 \beqn\label{eomsig}
  & (\partial^\mu\partial_\mu + m_\sigma^2)\sigma &= -g_\sigma\bar{\psi}\psi -g_2\sigma^2 -
  g_3\sigma^3,\\ \label{eomome}
  & \partial_\mu\Omega^{\mu\nu} + m_\omega^2\omega^\nu &=
    g_\omega\bar{\psi}\gamma^\nu\psi - c_3(\eta^\nu\omega^\nu)^3,\\ \label{eomrho}
  & \partial_\mu\vec{R}^{\mu\nu} + m_\rho^2\vec{\rho}^\nu &=
   g_\rho\bar{\psi}\gamma^\nu\vec{\tau}\psi +
   g_\rho\vec{\rho}_\mu\times\vec{R}^{\mu\nu},\\ \label{eompho}
  & \partial_\mu A^{\mu\nu} &=
   e\bar{\psi}\gamma^\nu\frac{1 - \tau_3}{2}\psi,
 \eeqn
for the mesons and photon, where the sum over all the particle
states in the no-sea approximation is adopted for the source term in
the Eqs.~(\ref{eomsig}, \ref{eomome}, \ref{eomrho}, \ref{eompho}).


\subsection{Relativistic Hartree-Bogoliubov Theory}

Usually, in the RMF theory, the mesons are treated as classical
fields. In order to describe the superfluidity of the nuclear
many-body system, one needs to quantize not only the nucleon but
also the meson fields. By using the well-known canonical
quantization method and the Green's function techniques, neglecting
retardation effects and the Fock term as it is mostly done in RMF,
one can get the so-called Relativistic Hartree-Bogoliubov (RHB)
equation,\cite{kucharek1991zpa}
 \beqn
 \label{dhb}
 \left(
    \begin{array}{cc}
        h - \lambda    &  \Delta  \\
        - \Delta^\ast  &  - h^\ast + \lambda
    \end{array}
\right) \left(
    \begin{array}{c}
        U_k  \\  V_k
    \end{array}
\right) &=& e_k\left(
    \begin{array}{c}
        U_k  \\  V_k
    \end{array}
\right),
 \eeqn
 where
 \beqn
 h &=& \bm{\alpha}\cdot\bm{p} + V + \beta(M + S),
 \eeqn
is the Dirac Hamiltonian with the scalar potential $S$ and vector
potential $V$,
 \beqn
 S &=& g_\sigma\sigma,~~~~ V ~=~
\beta\left(g_\omega\gamma^\mu\omega_\mu +
g_\rho\gamma^\mu\vec{\tau}\cdot\vec{\rho}_\mu
      + e\gamma^\mu\frac{1 - \tau_3}{2}A_\mu\right).
\eeqn

The pairing field is
 \beqn
   \Delta_{ab} &=& \frac{1}{2}
   \sum_{cd}\bar{V}_{abcd}\kappa_{cd},
 \eeqn
where $\bar{V}_{abcd}$ is the two-body effective interaction in the
particle-particle ($pp$) channel and the pairing tensor
$\kappa_{ab}=\sum_kV^\ast_{ak}U_{bk}$. The quasi-particle
eigenvectors are denoted as $(U_k, V_k)$, and $e_k$ is its
corresponding quasi-particle energies.

The chemical potential $\lambda$ in Eq.~(\ref{dhb}) is determined by
the particle number with the subsidiary condition, $\sum_k V^2_k ~=~
N$.


\subsection{Application to Symmetric Nuclear Matter}

For the static, uniform infinite nuclear matter, the coulomb field
is neglected, and the space-like components as well as the
differential of the time-like components of the meson fields vanish.
Furthermore, for symmetric nuclear matter, the $\rho$ meson has no
contribution on the mean field potential. Then the scalar potential
$S$ and vector potential $V$ are constants and have the simple form,
 \beqn
   S &=& g_\sigma \langle\sigma\rangle, ~~~~~~~ V ~=~ g_\omega \langle\omega_0\rangle.
 \eeqn
 The RHB equation (\ref{dhb}) can be decomposed into
($2\times2$) matrices of BCS-type,\cite{kucharek1991zpa}
 \beqn \label{dhb1} \left(
    \begin{array}{cc}
        \varepsilon(k) - \lambda  &  \Delta(k)                  \\
        \Delta(k)                 &  -\varepsilon(k) + \lambda
    \end{array}
\right) \left(
    \begin{array}{c}
        u(k)  \\  v(k)
    \end{array}
\right) &=& e(k)\left(
    \begin{array}{c}
        u(k)  \\  v(k)
    \end{array}
\right),
 \eeqn
 where the eigenvalue of the Dirac Hamiltonian for positive energies
 is denoted as, $\varepsilon(k)= V + \sqrt{k^2 + (m + S)^2}$,
 the Fermi energy $\lambda = \varepsilon(k_F)$,
and the quasi-particle energy $e(k) = \sqrt{(\varepsilon(k) -
\lambda)^2 + \Delta^2(k)}$. The corresponding occupation numbers
$v^2(k)$ have the form \beqn
   v^2(k)
&=&
   \frac{1}{2}
   \left(
      1 - \frac{\varepsilon(k) - \lambda}{\sqrt{[\varepsilon(k) - \lambda]^2 + \Delta(k)^2}}
   \right).
\eeqn

The pairing field $\Delta(k)$ obey the usual gap equation,
 \beqn
\label{deltak}
   \Delta(k)
&=&
   - \frac{1}{8\pi^2}\int_0^\infty f\cdot v_{pp}(k,p)\frac{\Delta(p)}{\sqrt{(\varepsilon(p)
   -\lambda)^2 + \Delta^2(p)}}p^2dp.
\eeqn
 where $f$ is an effective factor introduced to reduce the
{pairing potential}. The effective interaction in the $pp$ channel
$v_{pp}(k,p)$ is the one-meson exchange potential,
 \beqn
   v_{pp}(k,p) &=& v_{pp}^\sigma(k,p) + v_{pp}^\omega(k,p) + v_{pp}^\rho(k,p),
\eeqn
 where
  \beqn
   v_{pp}^\sigma(p,k)
&=&
   \frac{g_\sigma^2} {2\varepsilon^\ast(k) \varepsilon^\ast(p)}
    \left\{\frac{(\varepsilon^\ast(p) - \varepsilon^\ast(k))^2
   + m_\sigma^2 - 4m^{\ast2}}{4pk}
   \ln\frac{(k + p)^2 + m_\sigma^2}{(k - p)^2 + m_\sigma^2} - 1\right\},\\
   v_{pp}^\omega(p,k)
&=&
   \frac{g_\omega^2}{\varepsilon^\ast(k)\varepsilon^\ast(p)}\frac{2\varepsilon^\ast(k)\varepsilon^\ast(p) - m^{\ast2}}{2pk}
   \ln\frac{(k +p)^2 + m_\omega^2}{(k - p)^2 + m_\omega^2},\\
   v_{pp}^\rho(p,k)
&=&
   \frac{g_\rho^2}{\varepsilon^\ast(k) \varepsilon^\ast(p)}\frac{2\varepsilon^\ast(k)\varepsilon^\ast(p) - m^{\ast2}}{2pk}
   \ln\frac{(k +p)^2 + m_\rho^2}{(k - p)^2 + m_\rho^2},
 \eeqn
  with effective mass $m^\ast = m + g_\sigma\sigma$, and
$\varepsilon^\ast(k) = \sqrt{k^2 + m^{\ast2}}$.

The meson fields are replaced by their mean values, and could be
solved from the corresponding equations of motion by the various
giving nucleon densities,
 \beqn \label{eomsigma}
   m_\sigma^2\sigma
&=&
   - g_\sigma\rho_s - g_2\sigma^2 - g_3\sigma^3,\\
\label{eomomega}
   m_\omega^2\omega_0
&=&
   g_\omega\rho_v - c_3\omega_0^3,
\eeqn
 where $\rho_s$ and $\rho_v$ are respectively the scalar- and
baryon-density,
  \beqn
   \rho_s &=& \bar{\psi}\psi
          ~=~ \frac{2}{\pi^2}\int_0^{\infty}\frac{m+g_\sigma\sigma}{\sqrt{k^2 +
              (m+g_\sigma\sigma)^2}}v^2(k)k^2dk,\\
   \rho_v &=& \psi^\dag\psi
          ~=~ \frac{2}{\pi^2}\int_0^{\infty}v^2(k)k^2dk.
\eeqn

\section{Results and discussions}

For a given Fermi momentum $k_F$, the coupled Eqs.~(\ref{dhb1},
\ref{deltak}, \ref{eomsigma}, \ref{eomomega}) can be solved
self-consistently by iteration. The properties of $^1S_0$ pairing
correlations of symmetric nuclear matter are studied with the newly
developed effective interaction PK1 which takes into account the
self-interactions of $\sigma$-meson and $\omega$-meson as well as
the isospin dependence of the nuclear matter.\cite{longwh2004prc}

\begin{figure}[h]
 \centerline{\psfig{file=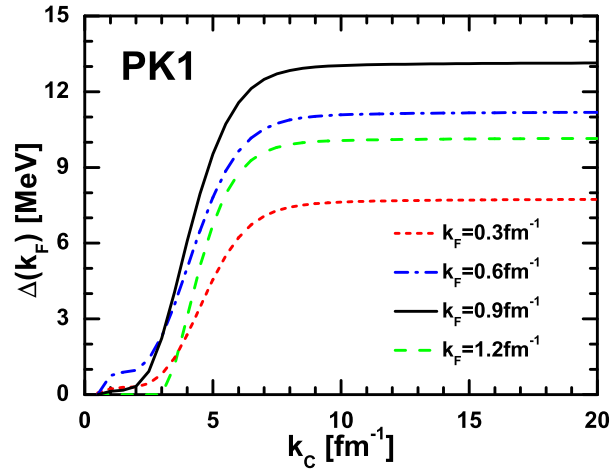,width=10cm}}
 \vspace*{8pt}
\caption{The pairing gap $\Delta(k_F)$ at the Fermi surface as a
function of the cutoff momentum $k_C$ in the momentum space for
different values of Fermi momentum $k_F$ = 0.3, 0.6, 0.9, and 1.2
fm$^{-1}$ with the effective interaction PK1.}
 \label{fig:fig1}
\end{figure}

The momentum integration in the gap equation, in principle, should
go to infinity. In actual calculations, it is necessary to have a
cutoff in the momentum space and the convergence of the pairing gap
on the cutoff momentum should be checked. The dependence of the
pairing gap on the cutoff momentum for different Fermi momentum is
given in Fig.~\ref{fig:fig1}. The results indicate that the cutoff
momentum $K_C \geq 10$ fm$^{-1}$ will guarantee the numerical
convergence. In the following, $K_C = 20$ fm$^{-1}$ will be adopted
and the corresponding effective interaction in the $pp$ channel, the
momentum-dependence of the pairing gap, and the influence of
effective interactions on the pairing gap at the Fermi surface,
etc., will be investigated.

\subsection{The effective interaction in the $pp$ channel}

\begin{figure}[h]
 \centerline{\psfig{file=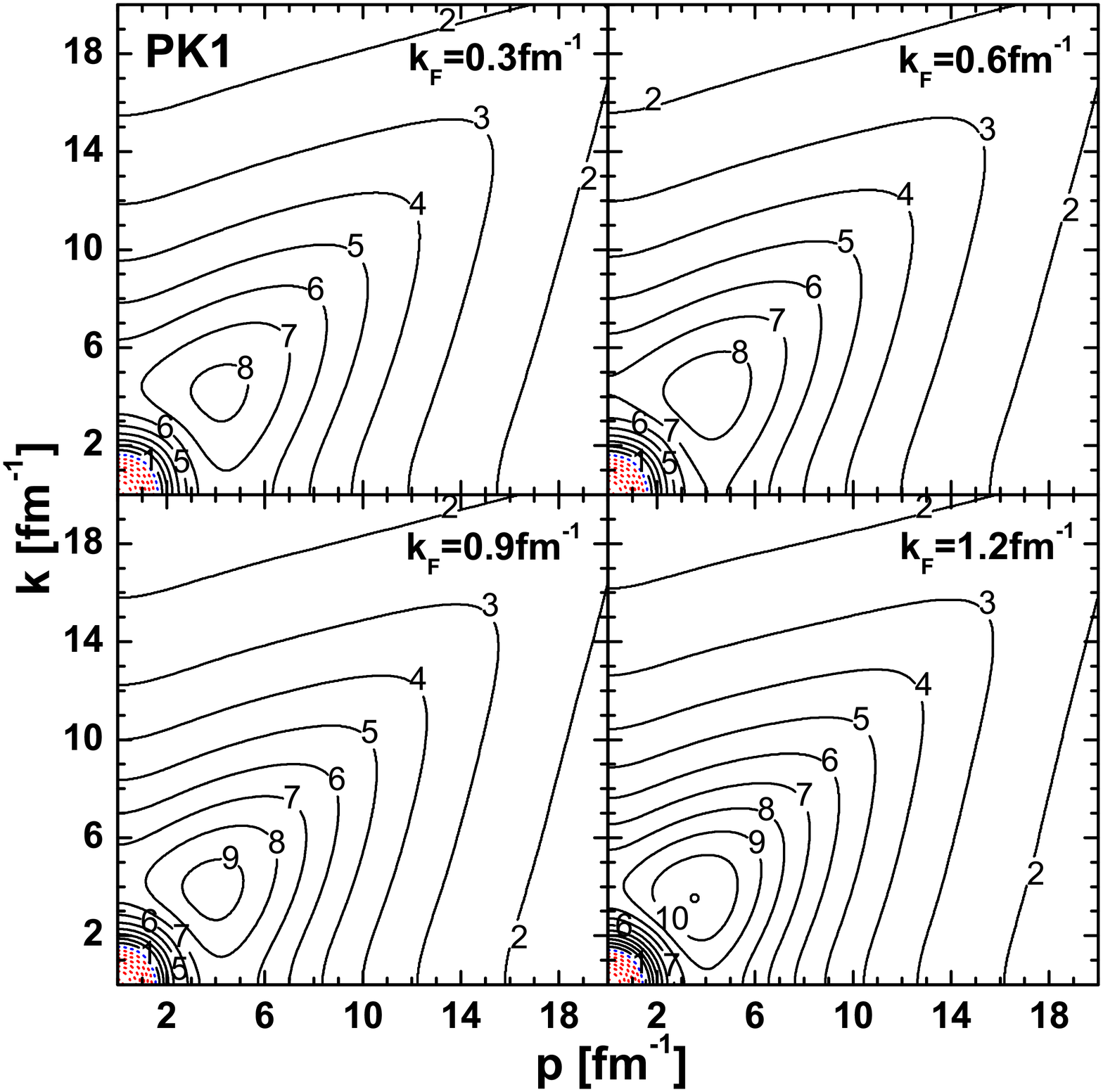,width=10cm}}
 \vspace*{8pt}
\caption{Contour plots for the  effective interaction in the $pp$
channel $v_{pp}(k,p)$ as a function of the momenta $p$ and $k$ for
different Fermi momentum $k_F$ = 0.3, 0.6, 0.9, and 1.2 fm$^{-1}$
with the effective interaction PK1. The contour lines have a
distance of 1 (fm$^2$) and the negative values are denoted as red
dashed lines.}
 \label{fig:fig2}
\end{figure}

The contour plot for the effective interaction in the $pp$ channel
$v_{pp}(k, p)$ for different Fermi momentum with the effective
interaction PK1 are shown in Fig.~\ref{fig:fig2}, where the contours
with negative values are denoted by red dashed lines.

The interaction is attractive for small and repulsive for larger
momenta $k$ and $p$, or equivalently, attractive for large distances
and repulsive for small distances. At around 1.5 fm$^{-1}$, the
interaction will change from being attractive to being repulsive.
The repulsive interaction reaches its maximum value at the momenta
$k$ and $p$ around 4 fm$^{-1}$. The maximum repulsive interaction
increases with the Fermi momentum.

\begin{figure}[h]
 \centerline{\psfig{file=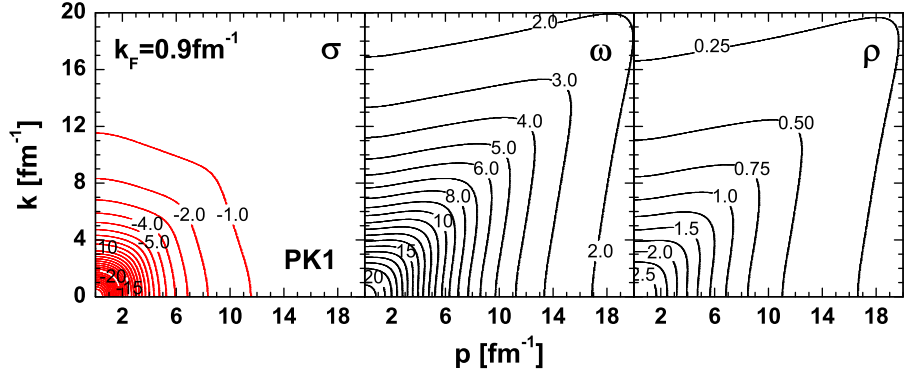,width=12cm}}
 \vspace*{8pt}
\caption{Contour lines of different mesons contributions for the
effective interaction in the $pp$ channel $v_{pp}(k,p)$ as a
function of the momenta $p$ for Fermi momentum $k_F$ = 0.9 fm$^{-1}$
with the effective interaction PK1. The contour lines have unit of
fm$^2$ and the negative contours are denoted as red colors.}
 \label{fig:fig3}
\end{figure}

The behavior of the effective interaction in the $pp$ channel
$v_{pp}(k, p)$ can be understood from contributions of different
mesons, as shown in Fig.~\ref{fig:fig3}. The scalar meson $\sigma$
provides the attractive part of the effective interaction,
$v_{pp}^\sigma(p,k)$, with a peak value at zero momentum, and
approaching zero with the increasing momentum. While the vector
meson $\omega$ and $\rho$ provide the repulsive part, which extend
to higher momenta than $v_{pp}^\sigma(p,k)$. The main contribution
for the repulsive part comes from the $\omega$ meson, as seen in the
figures, $v_{pp}^\omega(p,k)$ is one order of magnitude larger than
$v_{pp}^\rho(p,k)$.

\begin{figure}[h]
 \centerline{\psfig{file=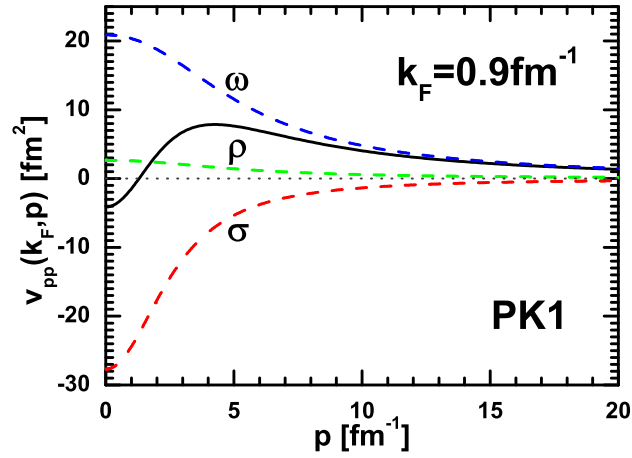,width=10cm}}
 \vspace*{8pt}
\caption{Different mesons contributions to the effective interaction
in the $pp$ channel $v_{pp}(k, p)$ as a function of $p$ at
$k=k_F=0.9~{\rm fm^{-1}}$ with the effective interaction PK1. The
dashed lines corresponds to that of the $\sigma$-, $\omega$- and
$\rho$-fields, the solid line represents the total contribution.}
 \label{fig:fig4}
\end{figure}

Different mesons contributions to the effective interaction in the
$pp$ channel $v_{pp}(k, p)$ as a function of $p$ at $k=k_F=0.9~{\rm
fm^{-1}}$ with the effective interaction PK1 are shown in
Fig.~\ref{fig:fig4}. The sum of all the mesons contributions has
considerable repulsive contributions for momenta larger than about
1.5 fm$^{-1}$, which is remarkably different from the one by the
Gogny force calculation.\cite{kucharek1991zpa}

\clearpage
\subsection{The momentum-dependence of the pairing gap}

\begin{figure}[h]
 \centerline{\psfig{file=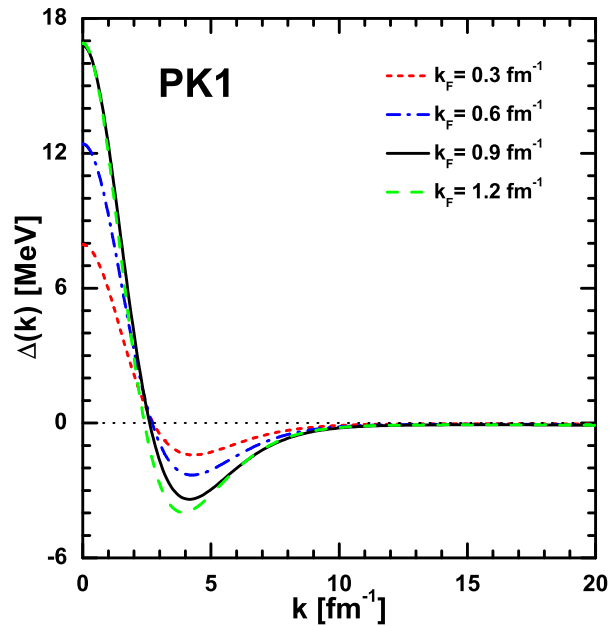,width=10cm}}
 \vspace*{8pt}
\caption{The pairing gap $\Delta(k)$ as a function of the momentum
$k$ for Fermi momentum $k_F$=0.3, 0.6, 0.9, and 1.2 fm$^{-1}$ with
the effective interaction PK1.}
 \label{fig:fig5}
\end{figure}

For a given $v_{pp}(k, p)$, the momentum-dependence of the pairing
gap $\Delta(k)$ can be obtained from Eq.~(\ref{deltak}). The
momentum-dependence of $\Delta(k)$ with different Fermi momentum is
shown in Fig.~\ref{fig:fig5}. The pairing gap has large and positive
values at small momenta, then decreases with the momentum and
changes its sign around 2.5 fm$^{-1}$. It continues to decrease till
$\sim$ 4.0 fm$^{-1}$, then slowly turns to zero. The difference of
pairing gap for different Fermi momentum is revealed mainly by the
behaviors at low momentum. At zero momentum, the largest pairing gap
occurs at the Fermi momentum 0.9 fm$^{-1}$.

\subsection{The pairing gap at the Fermi surface}

\begin{figure}[h]
 \centerline{\psfig{file=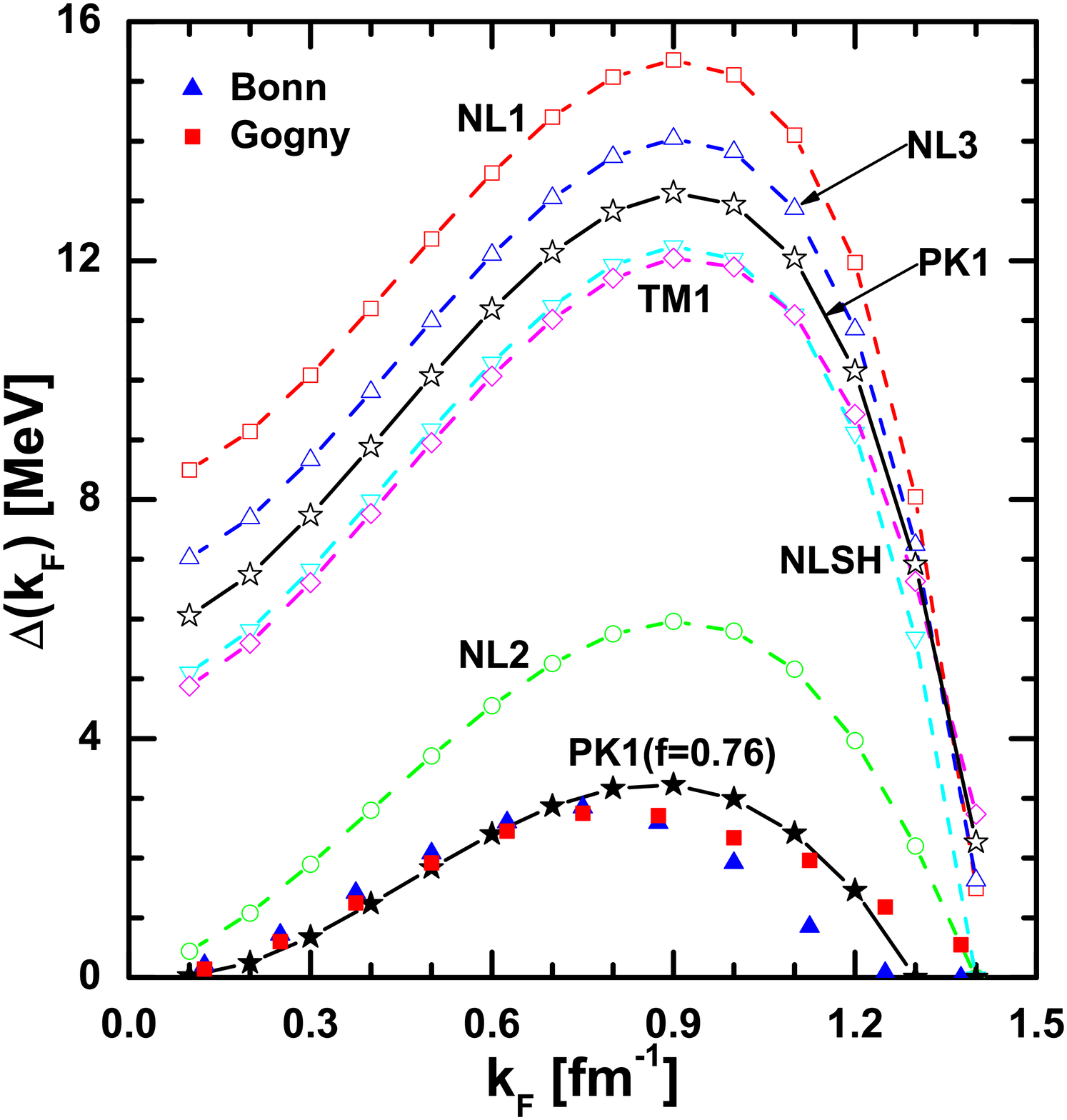,width=8cm}}
 \vspace*{8pt}
\caption{The pairing gap $\Delta(k_F)$ at the Fermi surface as a
function of the Fermi momentum $k_F$ for different effective
interactions. The results of Gogny and Bonn come from the
Ref.~[21].}
 \label{fig:fig6}
\end{figure}

One of most important properties of pairing gap is its value at the
Fermi surface. In Fig.~\ref{fig:fig6}, the pairing gap $\Delta(k_F)$
at the Fermi surface as a function of the Fermi momentum $k_F$ with
the effective interaction PK1 is shown in comparison with the
results obtained with the effective interactions NL1,\cite{NL1}
NL2,\cite{joon1986prl} NL3,\cite{lalazissis1997prc}
NLSH,\cite{sharma1993plb} TM1\cite{sugahara1994tp} and the results
calculated with Gogny force and Bonn potential.\cite{serra01}

It is found that the pairing gap $\Delta(k_F)$ is strongly dependent
on the nuclear matter density, or equivalently, the Fermi momentum.
The pairing gap $\Delta(k_F)$ increases with the Fermi momentum (or
density) and reaches maximum at Fermi momentum $k_F \approx 0.9$
fm$^{-1}$, then rapidly drops down to zero. Usually, the pairing gap
at the Fermi surface calculated with RMF effective interactions is
almost three times larger than the value calculated with Gogny
force. Moreover, the pairing gap at lower Fermi momentum does not
vanish in calculations with the RMF effective interactions, while it
does vanish in calculations with Gogny force or Bonn potential.

These differences come from the integral in Eq.~(\ref{deltak}) for
the pairing gap, which depends on the products of pairing gap
parameter $\Delta(p)$ and the effective interaction in the $pp$
channel $v_{pp}(k, p)$. From $v_{pp}(k, p)$ in Fig.~\ref{fig:fig2}
and $\Delta(p)$ in Fig.~\ref{fig:fig5}, it is found that
considerable contributions to the integral in Eq.~(\ref{deltak}) may
come from high momenta region. While the various effective forces in
RMF models are adjusted for mean-field calculations in Hartree
channel only, i.e., they are only valid for momenta below the Fermi
momentum\cite{kucharek1991zpa} and  the a realistic interaction in
the $pp$ channel $v_{pp}(k, p)$ can have very different behavior at
high momenta.

Considering that the RMF effective interactions give a too much
strong pairing field, an effective factor is introduced in the
particle-particle channel to reduce the pairing gap. For PK1, if a
factor $f=0.76$ is introduced, the resulting pairing gap is almost
the same as those with Gogny force or Bonn potential, and a maximum
pairing gap 3.2 MeV is obtained at Fermi momentum 0.9 fm$^{-1}$, as
shown in Fig.~\ref{fig:fig6}.

\clearpage
\section{Summary}

The pairing properties in the $^1S_0$ channel for symmetric nuclear
matter have been studied in the RMF theory with the effective
interaction PK1. The one-meson exchange potential is used in the
particle-particle channel in consistent with the particle-hole
channel. The effective interaction in the $pp$ channel is found to
be attractive at small momenta with attractive range around 1.5
fm$^{-1}$ and repulsive at large momenta. The pairing gap at the
Fermi surface is strongly dependent on the nuclear matter density.
It grows as the Fermi momentum increases, reaching its maximum
values at Fermi momentum around 0.9 fm$^{-1}$, then drops down to
zero rapidly. Considering the fact that the pairing gap at Fermi
momentum calculated with RMF effective interactions are three times
larger than that with Gogny force, an effective factor in the
particle-particle channel is introduced. For the effective
interaction PK1, a factor $f = 0.76$ will produce almost the same
results as those with Gogny force or with Bonn potential.

\section*{Acknowledgements}

This work is partly supported by Major State Basic Research
Development Program (2007CB815000) and the National Natural Science
Foundation of China (10435010, 10775004, and 10221003).


\end{document}